\documentclass[fleqn,12pt,twoside]{article}
\usepackage{espcrc1}

\usepackage{graphicx,epsfig}
\usepackage[figuresright]{rotating}

\newcommand{\AmS}{{\protect\the\textfont2
  A\kern-.1667em\lower.5ex\hbox{M}\kern-.125emS}}

\def\gsimeq{\,\,\raise0.14em\hbox{$>$}\kern-0.76em\lower0.28em\hbox  
{$\sim$}\,\,}  
\def\lsimeq{\,\,\raise0.14em\hbox{$<$}\kern-0.76em\lower0.28em\hbox  
{$\sim$}\,\,}  
\newcommand{\chem}[2]{$\rm{}^{#1}\kern-0.8pt#2$}
\newcommand{\chim}[2]{\rm{}^{#1}\kern-0.8pt#2}
\newcommand{\reac}[6]{$\rm\,{}^{#1}\kern-0.8pt{#2}\,({#3}\,,{#4})\,
           {}^{#5}\kern-0.8pt{#6}\,$}

\title{The r-process nucleosynthesis: a continued challenge for
nuclear physics and astrophysics}

\author{S. Goriely\address[IAA]{Institut d'Astronomie et
d'Astrophysique, ULB, CP 226, B-1050 Brussels, Belgium}
        \thanks{S.G. is FNRS research associate.}, 
       P. Demetriou\addressmark[IAA],
       H.-Th. Janka\address[MPA]{Max-Planck-Institut f\"ur Astrophysik, Postfach 1317,
            85741 Garching, Germany}
       J.M. Pearson\address[UM]{D\'ept. de Physique, Universit\'e de Montr\'eal,
Montr\'eal (Qc) H3C 3J7, Canada},  
  and M. Samyn\addressmark[IAA]
}
\begin{document}

\maketitle

\begin{abstract}
 The identification of the astrophysical site and the specific conditions in which
r-process nucleosynthesis takes place remain unsolved
mysteries of astrophysics. The present paper emphasizes some important future
challenges faced by nuclear physics in this problem, particularly in the determination
of the radiative neutron capture rates by exotic nuclei close to the neutron drip line
and the fission probabilities of heavy neutron-rich nuclei. These quantities are
particularly relevant to determine the composition of the matter resulting from the
decompression of initially cold neutron star matter. New detailed r-process calculations
are performed and the final composition of ejected inner and outer neutron star
crust material is estimated. We discuss the impact of the many uncertainties in
the astrophysics and nuclear physics on the final composition of the ejected 
matter. 
The similarity between the predicted and the solar abundance pattern for $A\ge 140$
nuclei as well as the robustness of the prediction with varied input parameters makes
this scenario one of the most promising that deserves further exploration.
\end{abstract}

\section{Introduction}

The rapid neutron-capture process, or r-process, is known to be
of fundamental importance for explaining the origin of approximately half
of the $A>60$ stable nuclei observed in nature. In recent years nuclear
astrophysicists have developed more and more sophisticated r-process models,
eagerly trying to add new astrophysical or nuclear physics ingredients to
explain the solar system composition in a satisfactory way. 
The r-process remains the most complex nucleosynthetic process to model from 
the astrophysics as well as nuclear-physics points of view. The site(s) of
the r-process is (are) not identified yet,  all the proposed scenarios facing serious
problems. Complex---and often exotic---sites have been considered in the hope of
discovering  astrophysical environments in which the production of neutrons is large
enough to give rise to a successful r-process. Progress in the modelling of type-II
supernovae and $\gamma$-ray bursts has raised a lot of excitement about the so-called
neutrino-driven wind model. However, until now no r-process can be simulated
{\it ab initio} without having to call for an arbitrary modification of the 
model parameters, leading quite often to physically unrealistic scenarios.

On top of the astrophysics uncertainties, the nuclear physics
of relevance for the r-process is far from being under control. The nuclear
properties of thousands of nuclei located between the  valley of $\beta$-stability and
the neutron drip line are required. These include the ($n,\gamma$) and ($\gamma, n$)
rates, $\alpha$- and $\beta$-decay half-lives, rates of $\beta$-delayed single and
multiple neutron emission, and the probabilities of neutron-induced, 
spontaneous, and $\beta$-delayed fission.  When considering complex
astrophysics sites like the neutrino-driven wind, proton-, $\alpha$-, and
neutrino-capture rates need to be estimated, too.

New developments of both nuclear physics and astrophysics aspects of the r-process
are discussed in this paper. Section 2 is devoted to the estimate
of the  neutron-capture rates and the fission probabilities of exotic neutron-rich
nuclei. These properties are of particular relevance for determining the still
poorly-known composition that results from the decompression of 
neutron-star (NS) crust matter, as shown in Sect.~3. 

\section{Neutron capture and fission by exotic n-rich nuclei}

 Although a great effort has been devoted in recent years to measuring decay
half-lives and reaction cross sections, the r-process involves so many (thousands)
unstable exotic nuclei for which so many different properties need to be known that only
theoretical predictions can fill the gaps. To fulfill these specific requirements, two
major features of nuclear theory must be contemplated, namely its {\it microscopic}
and {\it universal} aspects. A microscopic description by a physically sound model based
on first principles ensures a reliable extrapolation away from the experimentally known
region. On the other hand, a universal description  of all nuclear properties within one
unique framework for all nuclei involved ensures a coherent prediction of all unknown
data.   A special effort has been made recently to derive all the nuclear
ingredients of relevance in reaction theory from microscopic models \cite{go03}.  It
mainly concerns nuclear masses from a Hartree-Fock-Bogolyubov (HFB) model \cite{pea05},
nuclear level densities within the statistical model based on the microscopic
HF pairing strength and single-particle scheme \cite{dem01},
$\gamma$-ray strength functions from the HFB+QRPA approach \cite{khan04} and 
optical-model potentials from the Brueckner-HF approximation \cite{bau01}. Microscopic
estimates can lead to significant differences in comparison with more phenomenological
approaches.  These and their impact on the radiative neutron capture rate derived
within the Hauser-Feshbach statistical model are discussed in detail in
\cite{go03,go04} and are not repeated here.

\begin{figure}[t]
\begin{minipage}[t]{7.5cm}
\includegraphics[scale=0.31]{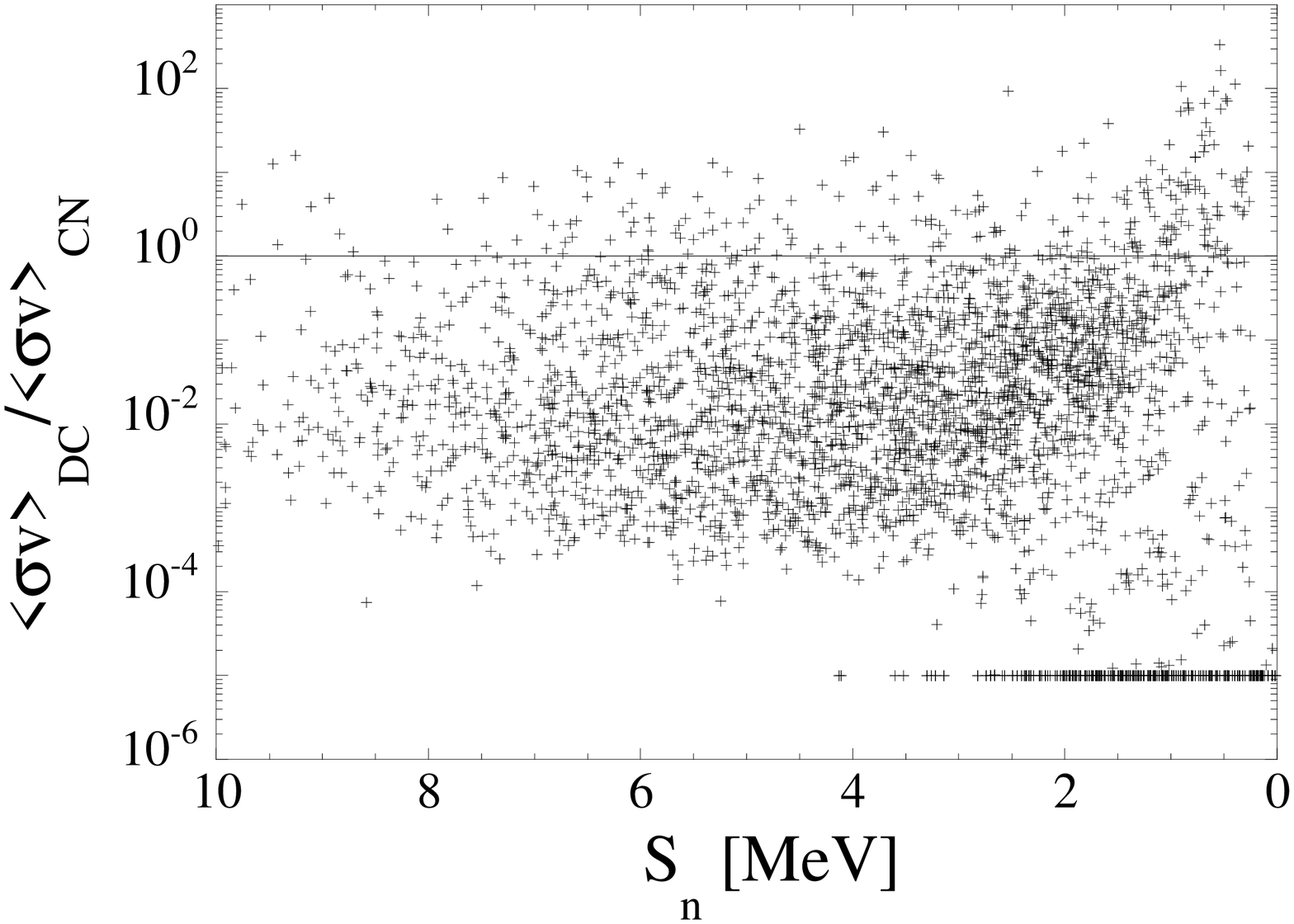}\\[-10mm]
\caption{Comparison of the DC and CN $(n,\gamma)$ rates for all nuclei with $20 \le Z \le
92$ located between the valley of stability and the neutron-drip line. A lower limit of
$10^{-5}$ is imposed.}
\label{fig_dc}
\end{minipage}
\hspace{\fill}
\begin{minipage}[t]{7.5cm}
\includegraphics[scale=0.28]{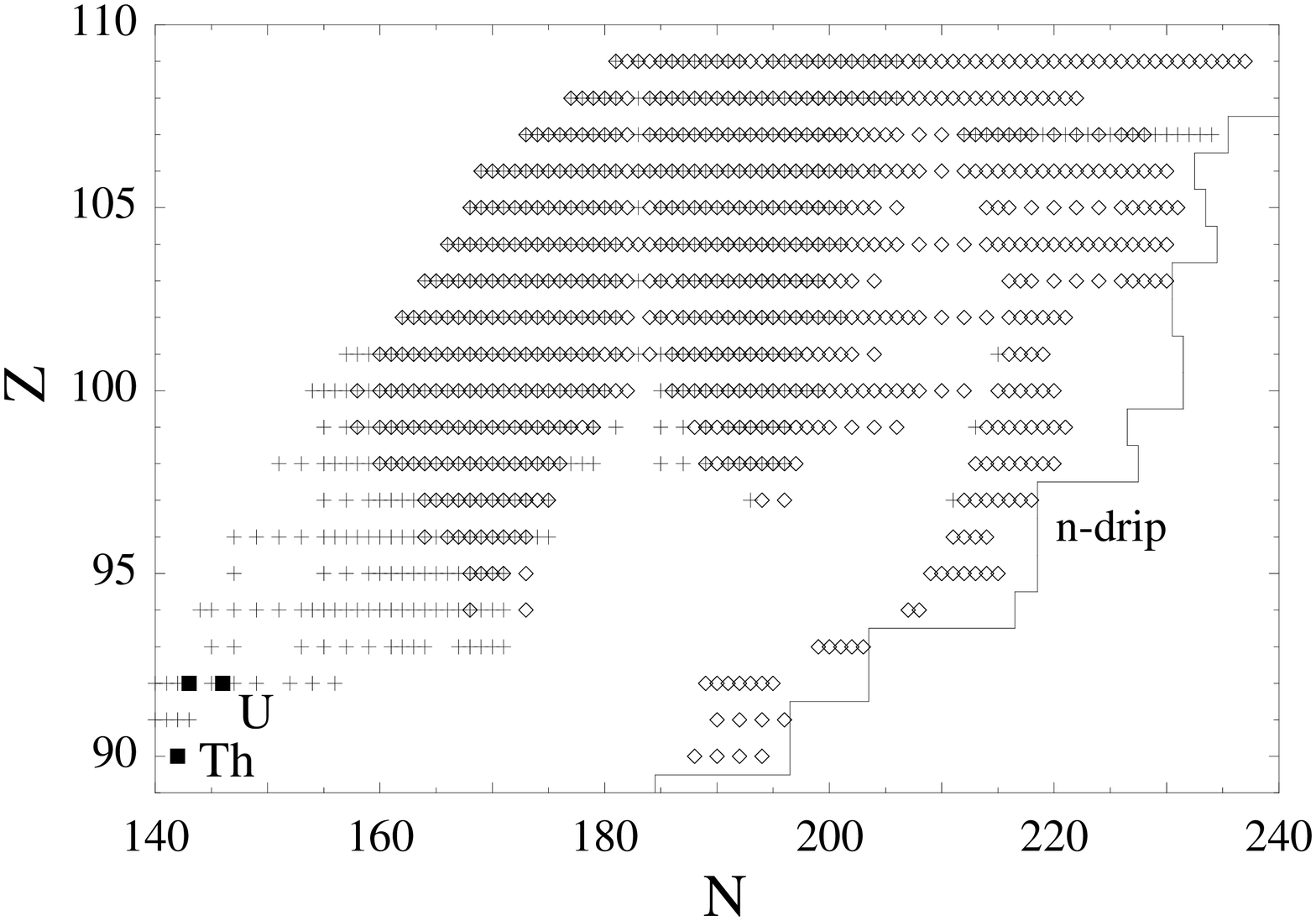}\\[-10mm]
\caption{Diamonds represent nuclei with a timescale against
spontaneous or $\beta$-delayed fission smaller than 1s, the (+)  those with
a n-induced fission rate larger than the radiative capture. }
\label{fig_fis}
\end{minipage}
\end{figure}

So far, all r-process calculations have made use of neutron capture rates evaluated
within the Hauser-Feshbach statistical model. Such a
model makes the fundamental assumption that the capture process takes place through the
intermediary formation of a compound nucleus (CN) in thermodynamic equilibrium. 
The formation of a compound nucleus is usually justified if the level density in the
compound nucleus at the projectile incident energy is large enough. However, when dealing
with exotic neutron-rich nuclei, the number of available states in the compound system
is relatively small and the validity of the Hauser-Feshbach model has to be questioned.
In this case, the neutron capture process might be dominated by direct electromagnetic
transitions to a bound final state rather than through the formation of a compound
nucleus. Direct captures (DC) are known to play an important role for light or closed
shell systems for which no resonant states are available.  The direct neutron capture
rates have been re-estimated for exotic neutron-rich nuclei as in \cite{go97} using a
modified version of the potential model to avoid the uncertainties affecting the
single-particle approach based on the one-neutron particle-hole configuration
\cite{go97}. It expresses the neutron DC cross section at an energy $E$ as
\begin{equation}
\sigma^{DC}(E) = \sum_{f=0}^x C_f^2 S_f \sigma_f^{DC}(E) 
+ \int_{E_x}^{S_n} \sum_{J_f,\pi_f} \langle C^2 S  
\rangle  \rho(E_f,J_f,\pi_f)\sigma_f^{DC}(E)dE_f 
\label{eq31}
\end{equation}

\noindent  where $x$ corresponds to the last experimentally known level
(in the final nucleus $f$) of excitation energy $E_x$ (smaller than 
the neutron binding energy $S_n$). Above $E_x$, the summation is replaced by a continuous
integration over the spin ($J$)- and parity ($\pi$)-dependent level
density $\rho$, and the spectroscopic factor $S$ and isospin
Clebsch-Gordan coefficient $C^2$ by an average quantity $\langle C^2 S
\rangle$. The DC cross section $\sigma_f^{DC}$ to each final state is
calculated within the potential model \cite{go97} in which the wave
functions of the initial and final systems are determined  by solving
the respective Schr\"odinger equations.
Because of the crucial sensitivity of the DC cross section to the spin
and parity assignment of the low-energy states in the residual
nucleus, only a microscopic combinatorial model of level density
is appropriate to the DC calculation.
New global calculations within the combinatorial method using the HFB single-particle
level scheme and $\delta$-pairing force are now available
\cite{hi01} and provide an accurate and reliable estimate of the intrinsic spin- and
parity-dependent level density. The average spectroscopic
factor remains very difficult to estimate. Experimental
systematics  suggest in a first approximation an energy-independent
value  $\langle C^2 S \rangle \simeq 0.06$ at energies lower than
$S_n$ \cite{go97}.  More
details on the sensitivity of the predicted DC rates to the
uncertainties affecting the nucleus-nucleon potential and the
spectroscopic factor can be found in
 \cite{go97}. The resulting DC rates are compared in Fig.~\ref{fig_dc} with the CN
rates for all the nuclei involved in the r-process nucleosynthesis. Interestingly,
the DC becomes larger than the CN contribution for many neutron drip nuclei. The lower
limit imposed on the DC/CN ratio (Fig.~\ref{fig_dc}) indicates that for some nuclei with
low $S_n$ the DC rates can become negligible, the selection rule forbidding the $E1$ or
$E2$ transitions to any of the available levels. Both the DC and CN mechanisms
may contribute to the radiative capture of a neutron. For this reason, the total capture
rate is often taken as the simple sum of both contributions, neglecting all possible
interferences. The statistical treatment inherent to the Hauser-Feshbach model might
however overestimate the resonant capture for the most exotic nuclei. The damping
procedure described in \cite{go97} is used here for the r-process
calculations described in Sect.~3. 

Another major difficulty in the nuclear modelling of the r-process concerns the
probabilities of the various fission processes, i.e., of spontaneous, 
$\beta$-delayed, and neutron-induced fission. 
Most of the r-process calculations either do not take into account fission processes at
all or only partially.  HFB calculations are presently in progress to improve the
predictions of fission barriers \cite{sam05}. In the meantime, we have used the
large-scale calculation of ETFSI fission barriers \cite{mam01} to re-estimate the
fission probabilities. Spontaneous and neutron-induced fission rates are determined
within the Hill-Wheeler and Hauser-Feshbach models, respectively, as described in
\cite{dem05}.  $\beta$-delayed fission rates are estimated considering the gross-theory
$\beta$-strength function \cite{ko75}.  Fig.~\ref{fig_fis} shows
the nuclear regions where the different fission modes influence the r-process flows.
Because of the strong ETFSI shell effect on the fission barriers of neutron-rich nuclei
around $N=184$, no fission recycling is expected before crossing
the $N=184$ closure. Spontaneous fission can affect the 
r-process nuclear flow quite substantially even close to the neutron drip line.
$\beta$-delayed fission is of small importance compared with the other
decaying modes.  Obviously, all the above conclusions should be taken with care, because
of the uncertainties remaining in the determination of the fission barriers and fission
probabilities. Future studies are needed, particularly in view of the importance of
the fission processes at specific r-process conditions, as described in Sect.~3.

\section{Decompression of initially cold NS matter}

The origin of the r-process nuclei is still a mystery. One of
the underlying difficulties is that the astrophysical site (and consequently the
astrophysical conditions) in which the r-process  takes place has not been identified.
Many scenarios have been proposed. The most favoured sites are all linked to 
core-collapse supernova or gamma-ray burst explosions. Mass ejection in the so-called
neutrino-driven wind from a nascent NS or in the prompt explosion of a supernova 
in the case of a small iron core or an O-Ne-Mg core have been shown to give rise to a
successful r-process provided the conditions in the ejecta are favorable with respect to
high wind entropies, short expansion timescales or low electron number fractions. 
Although these scenarios remain promising, especially in view of their
 significant contribution to the galactic enrichment \cite{argast03}, they remain
handicapped by large uncertainties associated mainly with the still incompletely
understood mechanism that is responsible for the  supernova explosion and the persistent
difficulties to obtain suitable r-process conditions in self-consistent dynamical
models. In addition, the composition of the ejected matter remains difficult to
ascertain due to the remarkable sensitivity of r-process nucleosynthesis to the
uncertain  properties of the ejecta.

Another candidate site has been proposed as possibly
contributing to the galactic enrichment in r-nuclei. It concerns the decompression of
initially cold NS matter \cite{lat77,meyer89,frei99}. In particular, special
attention has been paid to NS mergers due to their large neutron densities and the
confirmation by hydrodynamic simulations that a non-negligible amount of matter can be
ejected \cite{janka99,ros04}. 
Recent calculations of the galactic chemical evolution
\cite{argast03}, however, tend to rule out NS mergers as the dominant r-process site
for two major reasons: First, their relatively long life times and in particular
low rates of occurrence would lead to a sudden and
late r-process enrichment that is not compatible with the observed r-process enrichment 
of ultra-metal-poor stars. Second, the significant mass of r-process material ejected by
each event should lead to a large scatter of r-element overabundance in solar-like
metallicity stars; this scatter is not confirmed observationally. However, this
conclusion requires assumptions about the efficiency of mixing in the interstellar 
medium and is based on uncertain results from numerical models for the total amount 
of mass that is ejected by a single event. Moreover, it assumes a constant NS-NS merger
rate over the 10 Gyr galactic history that can be questioned. Another uncertainty
comes from the disregard of NS--black hole mergers that have been estimated to be about
10 times more frequent  than their NS-NS counterparts, although the possibility of mass
ejection is still unclear because of the unknown state of supranuclear matter in the 
NS~\cite{janka99,ros04}. But the ejection of initially cold, decompressed NS matter
might also happen in astrophysical scenarios like the explosion of a NS below its
minimum mass~\cite{sumi98} or in the spin-down phase of very rapidly rotating
supramassive or ultramassive NSs, which could lead to the equatorial shedding of
material with high angular momentum. The present study aims at providing a first
consistent calculation of the nucleosynthetic composition of dynamically ejected
material from cold NSs.  If the calculated abundances in the ejected matter turned out
not to be in agreement with solar-like r-abundances, this would tend to confirm the
growing opinion that the mergers of compact objects are unlikely to be important
contributors to the galactic r-enrichment.
A study of this kind is therefore of interest, filling an unfortunate
gap in the existing literature. 
In fact, little effort has so far been devoted to determining the 
composition of the matter that undergoes the decompression from initially cold
NS crust conditions. The first detailed calculation was
performed by Meyer~\cite{meyer89} in a systematic parametric study, but only included the
decompression down to densities around the neutron drip density $\rho_{\mathrm{drip}}\simeq
3~10^{11} {\rm g/cm^3}$. A second study came from \cite{frei99} 
but considered the
decompression below the drip density only. In this latter study, the composition of the
initial material was assumed to result from nuclear statistical equilibrium at high
temperatures ($T_9\simeq 6$), and its neutron-to-proton ratio was taken as a
free parameter. It is not plausible that such high temperatures can be reached in 
unshocked NS crust material that gets dynamically stripped from the NS surfaces during a 
merger and subsequently expands very quickly.
For this reason, we have performed new calculations to determine the final 
composition of clumps of initially cold NS inner and outer crust material after
decompression with a detailed treatment of the microphysics and thermodynamics during 
the decompression. All details can be found in \cite{go05}. 
The evolution of the matter density is modeled by considering the pressure-driven 
expansion of a self-gravitating clump of NS matter under the influence of tidal forces 
on an escape trajectory.
We characterize the expansion by defining the expansion timescale $\tau_{\mathrm{exp}}$
as  the time needed for the initial density to drop by three orders of magnitude.

As far as the outer crust is concerned, i.e., the NS material initially at a density below
$\rho_{\mathrm{drip}}$, the final isobaric abundance distribution after decompression is
essentially identical to the one prior to the ejection. Only $\beta$-decay (including
$\beta$-delayed neutron emission) can change the initial composition. We have redone
the calculation of \cite{ba71} with updated nuclear physics data to estimate the outer
crust composition. This calculation assumes
the matter to be in complete thermodynamic equilibrium and minimizes the free Gibbs
energy per cell to estimate the zero temperature composition. The energy of the
body-centered cubic lattice and relativistic electrons is included. For densities above
$ 3~10^9  {\rm g/cm^3}$, the matter is essentially  made of $N=50$ and $N=82$
r-process nuclei. More precisely, we find for
$10^9 \le \rho[{\rm g/cm^3}] \le 6~10^{10}$, $N=50$ nuclei with $80 \le A \le 86$. At
these densities, only nuclei with experimentally known masses are involved, so that this
result is free from theoretical uncertainties. This is not the case at larger densities
where we use the HFB-9 mass table \cite{pea05} to complement the compilation of
experimental masses. For $6~10^{10} \le \rho[{\rm g/cm^3}] \le 3~11^{10}$, $N=82$ nuclei
with $120
\le A \le 128$ populate the outer crust and should therefore enrich the
interstellar medium after ejection. The neutron emission by $\beta$-delayed
processes as well as the temperature effects on the energy distribution should spread
the matter over a wider mass range than the one originally found in the crust.

Regarding the inner crust, i.e initial densities $\rho>\rho_{drip}$, the situation is
quite different due to the existence of the neutron sea in which the nuclei
are immersed. The initial matter is assumed to be in $\beta$-equilibrium prior to the
expansion. This equilibrium is estimated on the basis of a Thomas-Fermi equation of
state (EOS)
\cite{onsi97} with the BSk9 Skyrme force used to build the HFB-9 mass table
\cite{pea05}. This leads at an initial density
$\rho\simeq 10^{14}  {\rm g/cm^3}$ to a composition characterized by the electron fraction
$Y_e=0.03$ corresponding to a cell made of a $Z=39$ and
$N=157$ nucleus. The expansion is followed down to the  neutron drip density as described
in \cite{meyer89}, allowing for the co-existence of Wigner-Seitz cells with different
proton numbers obtained through
$\beta$-transitions.
$\beta$-decays are estimated according to \cite{lat77} and found to heat the matter as
soon as $\mu_n-\mu_p-\mu_e>0$. When the matter reaches the neutron drip line, it is
distributed over a relatively large range of elements with $Z=40-70$, at least if
relatively fast $\beta$-transition probabilities and slow expansion
timescales are adopted (see \cite{go05} for more details). 

\begin{figure}[t]
\begin{minipage}[t]{7.5cm}
\includegraphics[scale=0.29]{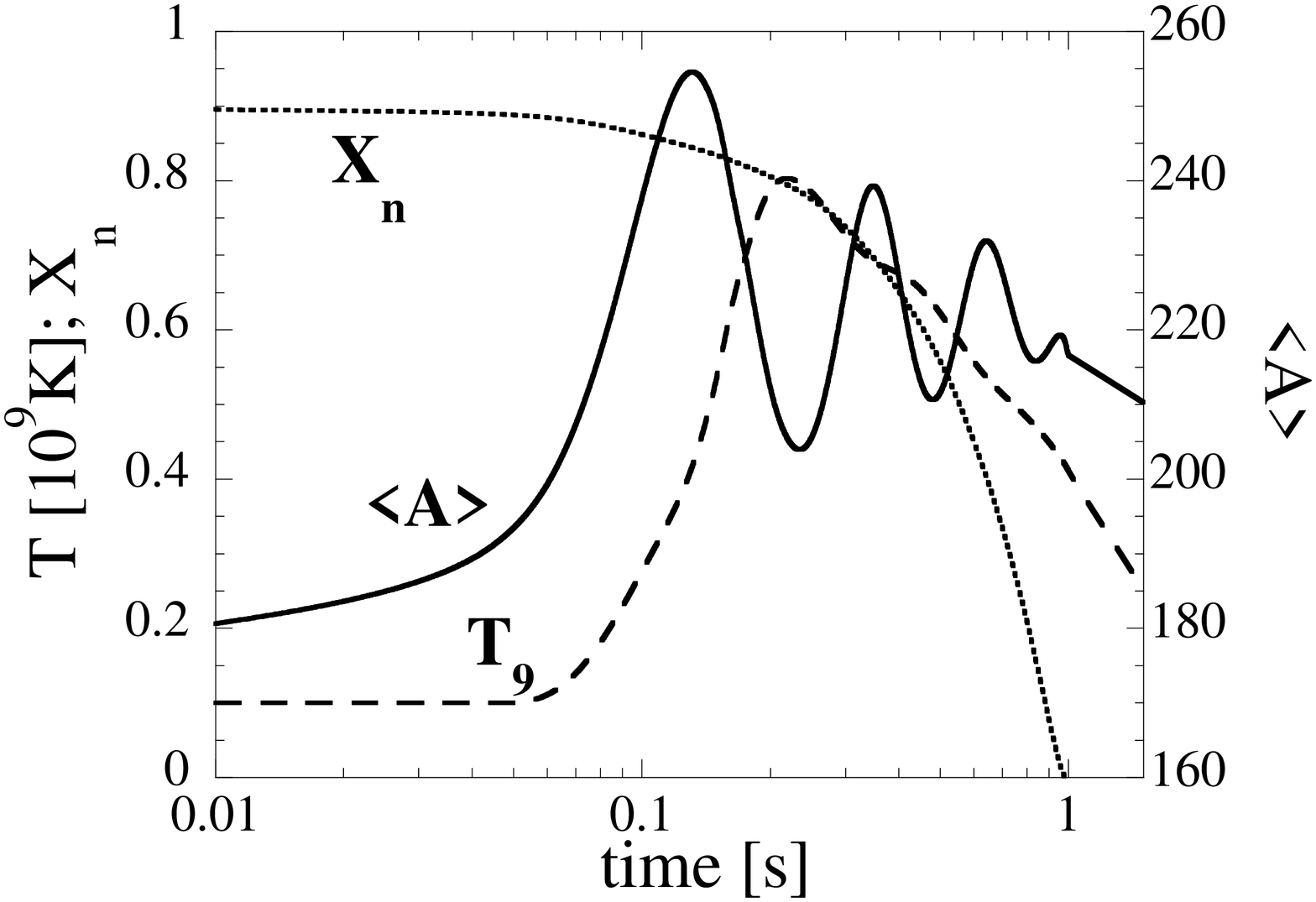}\\[-10mm]
\caption{Evolution of the temperature, $\langle
A \rangle$ and $X_n$ for a clump of material with initial density
$\rho = 10^{14}  {\rm g/cm^3}$, expanding with $\tau_{\mathrm{exp}}=6.5$~ms.
}
\label{fig_t9}
\end{minipage}
\hspace{\fill}
\begin{minipage}[t]{7.5cm}
\includegraphics[scale=0.29]{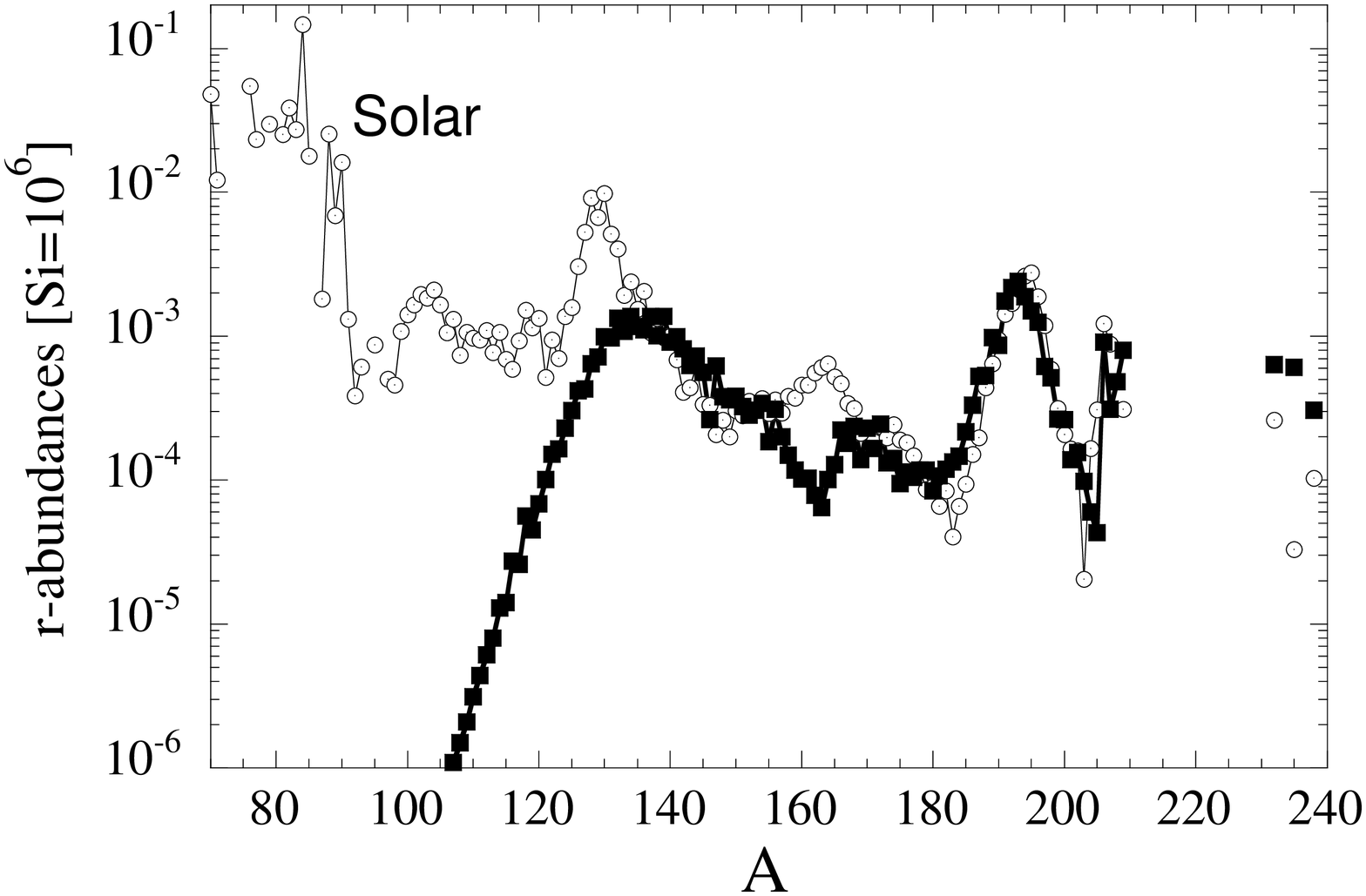}\\[-10mm]
\caption{Final r-abundance distribution for the same clump of material as in
Fig.~\ref{fig_t9}. The solar system distribution is also shown. }
\label{fig_ab}
\end{minipage}
\end{figure}

At the time the neutron drip density is reached, the matter is composed of drip
nuclei and of free neutrons at a typical density of $N_n\simeq 10^{35}~{\rm cm}^{-3}$.
The expansion is followed by a more ``traditional'' r-process reaction network including
the full description of radiative neutron captures, photodisintegrations, $\beta$-decays,
$\beta$-delayed neutron emissions as well as neutron-induced, $\beta$-delayed and
spontaneous fission processes. The rates for the 5000 nuclei involved are taken from the
microscopic calculations described in Sect.~2. The temperature increase is obtained with
the adequate EOS of \cite{tim99}, the nuclear
energy being supplied by the $\beta$-decays and fission reactions. In Fig.~\ref{fig_t9},
the time evolution of the temperature as well as the neutron mass fraction $X_n$ and
average mass number of heavy nuclei $\langle A \rangle$ is illustrated for a clump of
material with initial density $\rho = 10^{14}  {\rm g/cm^3}$, expanding on a timescale of
$\tau_{\mathrm{exp}}=6.5$~ms. The temperature does not exceed 
$T_9\approx 0.8$ and for such an expansion timescale all neutrons are captured. Fission
processes are found to recycle material leading to the oscillations of $\langle A
\rangle$ shown in Fig.~\ref{fig_t9}. The final abundance distribution obtained for such
conditions is displayed in Fig.~\ref{fig_ab}. The abundance distribution obtained for
$A>140$ is in relatively good agreement with the solar pattern. In particular the
$A=195$ peak is found at the right place with the right width. It should, however, be
stressed that such an r-abundance distribution results from a sequence of nuclear
mechanisms that significantly differ from those traditionally invoked
to explain the solar r-abundances, namely the establishment of an $(n,\gamma)-(\gamma,n)$
equilibrium followed by the $\beta$-decay of the corresponding waiting point. In the
present scenario, the neutron density is initially so high that the nuclear flow follows
for the first hundreds of ms after reaching the drip density a path touching the neutron
drip line. Fission keeps on recycling the material. After a few hundreds of ms,
the density has dropped by a few orders of magnitude and the neutron density experiences
a dramatic fall-off when neutrons get exhausted by captures (see Fig.~\ref{fig_t9}).
During this period of time, the nuclear flow around the $N=126$ region follows the
isotonic chain. When the neutron density reaches some $N_n=10^{20}$~cm$^{-3}$, the
timescale of neutron capture for the most abundant $N=126$ nuclei becomes larger than a
few seconds, and the nuclear flow is dominated by $\beta$-decays back to the stability
line. 

 In this scenario, photodisintegration reactions do not play any major role. An almost
identical abundance distribution is obtained when the temperature is kept at a constant
value of $T_9=0.1$. The distribution shown in Fig.~\ref{fig_ab} is found to be
robust to most of the nuclear uncertainties affecting masses, reaction rates and
$\beta$-decay rates. The fission fragment distribution adopted mainly affects the
abundance of the $A<140$ nuclei. The most sensitive parameter appears to be the expansion
timescale. As long as the expansion is relatively slow, all neutrons are captured and the
freeze-out takes place at densities around $N_n=10^{20}$~cm$^{-3}$ leading
systematically to an abundance distribution similar to the one shown in
Fig.~\ref{fig_ab}. In contrast, for fast expansions, i.e for $\tau_{\mathrm{exp}}<3$~ms in the
case of a clump at an initial density of $10^{14} {\rm g/cm^3}$, not all free neutrons are
captured, and the neutron density falls proportional to the density. In this case,
the final distribution becomes sensitive to the expansion timescale and can
differ significantly from the solar pattern. More details are given in \cite{go05}.

\section{Conclusion}
Most of the problems faced in understanding the origin of r-process elements
and observed r-abundances are related to our ignorance of the astrophysical site that is
capable of providing the required  large neutron flux. In this
respect, understanding the r-process nucleosynthesis is essentially an astrophysics
issue that will require improved hydrodynamic models to shed light on possible
scenarios. Different sites have been proposed, the currently most favoured ones 
being related to neutrino-driven outflow during supernova or $\gamma$-ray burst
explosions. Nevertheless, we have shown here that the decompression of initially cold NS
matter is also extremely promising. In particular, it provides suitable conditions for a
robust r-processing. The resulting r-abundance distribution is very similar to the solar
one, at least for $A>140$ nuclei. The underlying nuclear mechanisms, however, differ
significantly from  those acting in previous scenarios, in particular photoreactions do
not play a major role. The similarity between the predicted and solar abundance patterns
as well as the robustness of the prediction against variations of input parameters make
this site one of the most promising that deserves further exploration with respect to
various aspects such as nucleosynthesis, hydrodynamics and galactic chemical evolution.
From the nuclear point of view, this site also represents a new challenge, since it will
be necessary to determine
$\beta$-decay, neutron-capture and fission rates for nuclei which, in contrast to the
more conventional sites, are situated right at the neutron drip line.

\end{document}